\documentclass[conference]{IEEEtran}
\IEEEoverridecommandlockouts
\usepackage{cite}
\usepackage{amsmath,amssymb,amsfonts}
\usepackage{amsthm}
{
	\newtheorem{prob}{Problem}
	\newtheorem{corollary}{Corollary}
	\newtheorem{proposition}{Proposition}
	\newtheorem{definition}{Definition}
	\newtheorem{lemma}{Lemma}
	\newtheorem{theorem}{Theorem}
	
}
\usepackage{algorithmic}
\usepackage{graphicx}
\usepackage{textcomp}
\usepackage{xcolor}
\usepackage{enumerate}
\usepackage[margin=0.75in]{geometry}
\def\BibTeX{{\rm B\kern-.05em{\sc i\kern-.025em b}\kern-.08em
    T\kern-.1667em\lower.7ex\hbox{E}\kern-.125emX}}
\begin{document}

\title{\huge{Minimizing Age of Information for Real-Time Monitoring in Resource-Constrained Industrial IoT Networks%Average Age of Information Minimization for Resource-Constrained  IoT Networks%: An MDP Approach %Timely Status Update in Resource-Constrained IoT Systems with Random Arrival: An MDP Approach
}}

%\author{\IEEEauthorblockN{Qian Wang, He Chen, Yonghui Li,  Branka Vucetic}%
%	\IEEEauthorblockA{School of Electrical and Information Engineering, The University of Sydney, Sydney, Australia} 
%	%\{qian.wang2, he.chen, yonghui.li, branka.vucetic\}@sydney.edu.au}%
%%	\IEEEauthorblockA{\IEEEauthorrefmark{2}\textit{School of Electronic Information and Communications}, Huazhong University of Science and Technology, Wuhan, China\\
%		%City, Country \\
%		\{qian.wang2, he.chen, yonghui.li, branka.vucetic\}@sydney.edu.au, %\IEEEauthorrefmark{2}xhge@hust.edu.cn}
\author{\IEEEauthorblockN{Qian Wang\IEEEauthorrefmark{1}, He Chen\IEEEauthorrefmark{1}, Yonghui Li\IEEEauthorrefmark{1}, Zhibo Pang\IEEEauthorrefmark{2}, Branka Vucetic\IEEEauthorrefmark{1}}%
	\IEEEauthorblockA{\IEEEauthorrefmark{1}School of Electrical and Information Engineering, The University of Sydney, Sydney, Australia} 
	%\{qian.wang2, he.chen, yonghui.li, branka.vucetic\}@sydney.edu.au}%
	\IEEEauthorblockA{\IEEEauthorrefmark{2}Automation Solutions, ABB Corporate Research, V\"aster\aa  s, Sweden, \\
		\IEEEauthorrefmark{1}\{qian.wang2, he.chen, yonghui.li, branka.vucetic\}@sydney.edu.au, \IEEEauthorrefmark{2}pang.zhibo@se.abb.com}
%\and
%\and
%\and
}

\maketitle

\begin{abstract}
This paper considers an Industrial Internet of Thing (IIoT) system with a source monitoring a dynamic process with randomly generated status updates. The status updates are sent to an designated destination in a real-time manner over an unreliable link. The source is subject to a practical constraint of limited average transmission power. Thus, the system should carefully schedule when to transmit a fresh status update or retransmit the stale one. To characterize the performance of timely status update, we adopt a recent concept, \textit{Age of Information} (AoI), as the performance metric. We aim to minimize the long-term average AoI under the limited average transmission power at the source, by formulating a constrained Markov Decision Process (CMDP) problem. To address the formulated CMDP, we recast it into an unconstrained Markov Decision Process (MDP) through Lagrangian relaxation. We prove the existence of optimal stationary policy of the original CMDP, which is a randomized mixture of two deterministic stationary policies of the unconstrained MDP. We also explore the characteristics of the problem to reduce the action space of each state to significantly reduce the computation complexity. We further prove the threshold structure of the optimal deterministic policy for the unconstrained MDP. Simulation results show the proposed optimal policy achieves lower average AoI compared with random policy, especially when the system suffers from stricter resource constraint. Besides, the influence of status generation probability and transmission failure rate on optimal policy and the resultant average AoI as well as the impact of average transmission power on the minimal average AoI are unveiled.
\end{abstract}

%\begin{IEEEkeywords}
%component, formatting, style, styling, insert
%\end{IEEEkeywords}

\section{Introduction}

% 1.IoT introduction; 2. Special IoT requirement for AoI, AoI description 3. our work with AoI meaning; 4 result.
%The IoT has a wide range of applications, ranging from individual to industry, such as health monitoring, smart building, smart city and so on \cite{ib2}. Actual system requirement varies from different IoT application scenarios. 
The Internet of Thing (IoT) aims to connect a massive number of devices with different objectives and functions so as to bring an unprecedented information network and achieve value increment \cite{ib1}. The application of IoT technologies in industrial environment is normally referred to as Industrial Internet of Thing (IIoT)\cite{io1}, which provides pervasive connectivity to sensors, actuators and controllers in Industrial Control Systems (ICS). Real-time monitoring is pivotal for IIoT, especially for manufacturing process in industrial automation, where the controller needs to make sure every equipment is under precise control. Moreover, as the first step of network intrusion detection, real-time monitoring also plays a critical role in securing the ICS \cite{is1}. In real-time monitoring, the timely delivery of system status updates from IIoT devices to the controllers is essential. As timely update of system status monitored by IIoT devices is fundamentally different from the conventional throughput maximization and delay minimization problems, a novel performance metric \textit{Age of Information} (AoI) has been introduced in \cite{ib3}. It is defined as the time elapsed since the generation time of latest received status at the destination.

Recent years have witnessed considerable efforts on analyzing the AoI of various systems \cite{ib3,ib12,yfan} and exploring the optimal sampling and updating policies to minimize system AoI \cite{yfan,ib6,ib9,ib4,ib7,b0}. The average AoI of systems modeled by first-come-first-served (FCFS) and last-come-first-served (LCFS) queues were analyzed in \cite{ib3} for both $M/M/1$ and $D/M/1$ queuing models, and in \cite{ib12} for $M/M/1$ model, respectively. In \cite{ib3,ib12}, both the cases with and without packet preemption were considered, while \cite{ib3} further optimized the status generation rate at the source to minimize system average AoI. %in \cite{ib3}. 
%Recent years have witnessed considerable efforts on analyzing the AoI of various systems \cite{ib3,ib12,ib10,ib13} and exploring the optimal sampling and updating policies to minimize system AoI \cite{ib6,ib9,ib4,ib7,b0}. The average AoI of systems modeled by first-come-first-served (FCFS) and last-come-first-served (LCFS) queues were analyzed in \cite{ib3} for both $M/M/1$ and $D/M/1$ queuing models, and in \cite{ib12} for $M/M/1$ model, respectively. In \cite{ib3,ib12}, both the cases with and without packet preemption were considered and the status generation rate at the source was further optimized to minimize system average AoI in \cite{ib3}. Peak AoI (PAoI) was first defined and studied in \cite{ib13,ib10}. In \cite{ib13}, the PAoI with packet management was investigated. The PAoI of $M/M/1$ queuing systems with packet loss was analyzed in \cite{ib10}, wherein the average peak AoI was derived for LCFS queue with packet retransmission.

Recent work on the optimization of the AoI of various systems can be grouped into two categories considering either randomly generated (arrived) status update model \cite{yfan,ib6,ib9} or \textit{generate at will} status update model \cite{ib4,b0}. Queue management for systems with multiple sources that randomly generate status and share one common transmitter was investigated in \cite{ib6}. Comparing to $M/M/1$ queues, single queue technique in \cite{ib6} reduces transmissions and achieves lower average AoI. The single source system with randomly generated status according to a Bernouli process was considered in \cite{ib9}. In the system, the transmission of each status update is assumed to take a fixed number of time slots and suffer from no error. The corresponding optimal transmission schedule to decide whether to skip or switch to transmit a new generated status was determined. The optimal status update policy for a status monitoring system with \textit{generate at will} model was studied in\cite{ib4} and was shown to be superior to the zero-wait policy in many scenarios. 

In practice, IIoT devices are normally energy-constrained. As such, there is a strong demand for energy efficient policies and  techniques. Rather than simply focusing on AoI minimization, power limitation of IIoT devices has been recently considered in system designs. Tradeoff between AoI and energy consumption has been derived in \cite{yfan} as well as the optimal transmission policy to minimize average AoI. Energy harvesting techniques with finite battery capacity were considered in \cite{ib7} for AoI minimization, where the optimal transmission policy to minimize the long-term average AoI is proved to be a renewal policy. However, the channel model of \cite{ib7} was assumed to be error free as in \cite{ib9}, which is impractical in real applications. A very recent  work \cite{b0} considered the error prone channel model and developed the optimal status update policy to minimize average AoI while taking resource constraints into account for the \textit{generate at will} model. A key conclusion obtained in \cite{b0} is that, when the resource is limited, not all status updates should be transmitted. This leads to a natural question: when the status update is randomly generated, how will the status generation probability influence the long-term average AoI under the constraint of limited resource? To the best knowledge of the authors, this is still an open question.

%Motivated by the gap, we consider an IoT system with a source that monitors dynamics with randomly generated status updates and sends the status to its destination over an unreliable link. Under a practical constraint of limited
%average transmission power for the IoT device, we develop an optimal transmission scheduling policy to minimize long term average AoI. Considering the error prone channel and randomly generated status update, in terms of limited average transmission power, the system needs to carefully decide when to transmit fresh status update and do retransmission. The coupled transmission failure and status generation makes this problem challenging as instantaneous AoI drops only when status transmission is successful. It is the uncertainty of status generation and transmission failure that result in uncertain AoI variation. To solve the formulated problem, we recast it into a constrained Markov Decision Process problem. We first prove the existence of optimal stationary policy which is a randomized mixture of two deterministic stationary monotone policies. Finally, we conduct simulations to illustrate the influence of status generation rate and transmission failure rate on optimal policy and the resultant average AoI as well as impact of average transmission power on optimal average AoI.
Motivated by the gap, we consider an Industrial IoT system with an IIoT device (source) that monitors a dynamic process with randomly generated status updates and sends the status to its destination (e.g., controller) over an unreliable link. Under a practical constraint of limited average transmission power for the IIoT device, we develop an optimal transmission scheduling policy to minimize the long-term average AoI. Considering the  limited average transmission power at the source, the system needs to carefully decide whether to transmit a fresh status update or retransmit the stale one at the beginning of each time slot. Note that the AoI is jointly affected by the transmission failure and status generation process, and the instantaneous AoI drops only when a status update is transmitted successfully, which makes the considered problem challenging. In particular, the uncertainty of status generation and transmission failure result in uncertain AoI variation. To find the optimal decision policy, we formulate the considered problem into a constrained Markov Decision Process (CMDP) problem and transform it into an unconstrained Markov Decision Process (MDP) through Lagrangian relaxation. We prove the existence of optimal stationary policy for the original CMDP, which is a randomized mixture of two deterministic stationary policies of the unconstrained MDP. We also explore the characteristics of our CMDP problem to reduce the action space of each state so as to reduce the computation complexity. We further prove the threshold structure of the optimal deterministic policy for the unconstrained MDP. Thanks to the identified threshold structure, only action shifting boundary is needed, hence the required memory at the IIoT device to execute the policy is reduced. Finally, simulation results are provided, which show that the proposed optimal policy achieves lower average AoI compared with random policy, especially when the system suffers from stricter resource constraint. Besides, the influence of status generation probability and transmission failure rate on optimal policy and the resultant average AoI, as well as the impact of average transmission power on the minimal average AoI are illustrated through numerical results.
% illustrate the influence of status generation probability and transmission failure rate on optimal policy and the resultant average AoI as well as the impact of average transmission power on the minimal average AoI. The proposed optimal policy achieves superior performance with lower average AoI comparing with random policy especially when suffering serious resource constraint.
%we conduct simulations to verify the superior performance of proposed optimal policy which has lower average AoI compared with random policy, especially when the system suffers from stricter resource constraint.

\section{System Model and Problem Formulation}
\subsection{System model}

We  consider a discrete-time IIoT monitoring system where a single IIoT device (e.g., sensors) monitors a dynamic process and transmits status update to a destination (e.g., controller) through an unreliable link with a constant transmission power. At the beginning of time slot $t$, the IIoT device randomly generates a status update according to an independent and identically distributed (i.i.d) Bernoulli process $B(t)$, with parameter $p$ \cite{ib9}, and needs to decide whether to transmit the fresh status update or perform a retransmission of the previously unsuccessful status update. Successful reception of a state update is acknowledged by the feedback signal (ACK/NACK) from destination to the IIoT device, which is assumed to be transmitted through perfect channel (error free and delay free) \cite{b0}.  
%Moreover, we consider the IoT device has no buffer to store status update. As such, once the IoT device generates new update, it needs to decides whether to transmit fresh status update or drop the new arrival. When there is no fresh data generated, the devices should decides whether continues to transmit stale status.
We consider that it takes constant time to transmit an update from IIoT device to destination, which is assumed to be equal to the duration of one time slot for simplicity. There is no buffer at the IIoT device. Hence, once a new update is generated, the IIoT device needs to decide whether to transmit or drop the new status update. Besides, when status transmission failure occurs over the error prone link, the IIoT device needs to make another decision on whether to retransmit the current status or not. We define the set of the IIoT device actions as $A = \{1,2,3\} $. At the beginning of each slot, the IIoT device needs to choose one action $a_t \in A$, if $a_t=1$, the IIoT device does not transmit any status update; if $a_t=2$, the IIoT device retransmits previously failed update; otherwise, the IIoT device transmits a new update. Following classical Automatic Repeat reQuest (ARQ) protocol, we assume that status update transmission failures of different time slot are independent and not relevant to transmission attempts. The following equality holds for the considered transmission model,
\begin{equation}
\label{e0}
P(failure|a(t)\neq 1)=\gamma.
\end{equation}
%\begin{assumption}
%	\label{as:1}
%	The packet transmission failures of different time slot are independent
%\end{assumption}

The AoI measures data freshness at the receiver, defined as time elapsed since the latest successfully received status update was generated\cite{ib3}. Denote by $u(t)$ the generation time for latest received update, the AoI of the IIoT device at destination $\triangle(t)$ is defined as the random process, 
\begin{equation}
\label{e1}
\triangle(t)=t-u(t).
\end{equation} Hence, the AoI decreases to the total transmission time of a status update when it is received and successfully decoded.

Now, we define the state of the system at time $t$ as $s_t$. Specifically, $s_t=(\delta_t,l_t,b_t)$, where $\delta_t \in \{1,2,3,...\}$ denotes the AoI at the beginning of time slot $t$. $l_t \in \{0,1,2,..., l_{max}\}$ denotes the total transmission times of last transmitted status update at the beginning of time slot $t$, $l_t=0$ when there is no status update being transmitted at previous time slot, and $l_{max}$ is the maximum allowable transmission times of each status update. And $b_t \in \{0,1\}$ indicates whether a new status update is generated by the IIoT device: $b_t=1$ denotes the generation of a new update at the beginning of time slot $t$, $b_t=0$, otherwise. We have $\delta_t \geq l_t$, according to the definition of AoI in \eqref{e1}.

Status update is retransmitted when NACK is received at the IIoT device in classical ARQ protocol. However, as for the AoI framework, it is meaningless to retransmit failed outdated information when a fresh status update is available since the transmissions of the new and outdated statuses will suffer from the same transmission failure probability. Moreover, for the IIoT device, limited power leads to restricted transmission. It is a waste of energy to keep transmitting failed out-of-date status update when no fresh status update is generated. Because, anyhow, the average transmission power is limited. The AoI will not be considerably decreasing after a large number of retransmissions. On the other hand, new status update might be generated in the next slot and transmitting a fresh status update can lead to significant AoI drop. In addition, frequently updating instant status update helps to lower the AoI at the cost of higher average transmission power. Consequently, we impose an average transmission power constraint at the IIoT device. As it is assumed that the IIoT device uses a constant transmission power, average transmission power constraint is the same as a constraint on average transmission probability, denoted by $\Gamma_{max} \in (0,1]$.

At the beginning of each time slot, the IIoT device determines whether to transmit a new status update, retransmit failed stale status or remain idle to minimize the average AoI under average transmission power constraint. Define $\pi$ as a stationary scheduling policy, that maps the state $s$ to action $a$ neither deterministically or probabilistically. Let $\Pi$ be the set of all feasible policies. The objective of AoI minimization with limited average power can be formulated as a constrained Markov Decision Process (CMDP)\cite{b1}, described by a tuple $\{S,A,P,r,d\}$, where 
    \begin{itemize}
	\item The countable state space $S=\{(\delta,l,b)\}$ is as above,
	\item The action space $A=\{1,2,3\}$ is already defined above,
	\item $P$ are the transition probabilities, $P(s_{t+1}|s_t,a_t)$ is the probability of moving from state $s_t$ to $s_{t+1}$ when taking action $a_t$. Under the considered i.i.d. status generation model and transmission model at each slot, we have
	\begin{small}
	   \begin{equation}
	   \label{e2}
		\begin{aligned}
		P(s_{t+1}|s_t,a_t)&=P( \delta_{t+1},l_{t+1},b_{t+1}|\delta_t,l_t,b_t,a_t)\\
		&=P(\delta_{t+1},l_{t+1}|\delta_t,l_t,b_t,a_t)P(b_{t+1}).\\
		%	      	P(\delta_t+1,l_t+1,1|\delta_t,l_t,b_t,a_t=1)=\gamma p_0\\
		\end{aligned}
		\end{equation}
	\end{small} To be more specific,
		\begin{small}
			\begin{equation}
			\label{e3}
			\begin{aligned}
			P(\delta_t+1,l_t+1,b_{t+1}|\delta_t,l_t,b_t,a_t=2)&=\gamma P(b_{t+1}),\\
			P(l_t+1,l_t+1,b_{t+1}|\delta_t,l_t,b_t,a_t=2)&=(1-\gamma) P(b_{t+1}),\\
			P(\delta_t+1,1,b_{t+1}|\delta_t,l_t,b_t=1,a_t=3)&=\gamma P(b_{t+1}),\\
			P(1,1,b_{t+1}|\delta_t,l_t,b_t=1,a_t=3)&=(1-\gamma) P(b_{t+1}),\\	
			P(\delta_t+1,0,b_{t+1}|\delta_t,l_t,b_t,a_t=1)&= P(b_{t+1})	,    
			\end{aligned}
			\end{equation}
		\end{small} and otherwise, $P(s_{t+1}|s_t,a_t)=0$,
	\item  $r: S \times A  \rightarrow R$ is the immediate reward with the reward function of state-action pairs being defined as $r((\delta,l,b),a)=\delta$,
	\item $d: S \times A \rightarrow R$ is the immediate costs taking action $a$ in state $s$. Cost function of state-action pairs is
	\begin{small}
   $d((\delta,l,b),a)=
	\left\{
	\begin{array}{rcl}
	0,& &\text{if }  a=1 \\
	1,& &\text{otherwise. }  \\
	\end{array}
	\right.$
	\end{small}
\end{itemize}

Given initial state $s_0$, the infinite-horizon average reward of any feasible policy $\pi \in \Pi$ is 
\begin{small}
\begin{equation}
\label{e4}
      C(\pi|s_0)=\lim_{T \rightarrow \infty}\sup\frac{1}{T} \sum_{k=0}^{T}{\mathbb E}_{s_0}^\pi[r(s_k,a_k)|s_0].
\end{equation} 
\end{small}Define the infinite-horizon average cost with respect to
policy $\pi \in \Pi$ as 
\begin{small}
\begin{equation}
\label{e5}
D(\pi|s_0)=\lim_{T \rightarrow \infty}\sup\frac{1}{T} \sum_{k=0}^{T}{\mathbb E}_{s_0}^\pi[d(s_k,a_k)|s_0].
\end{equation}
\end{small}
Here, ${\mathbb E}[\cdot]$ denotes the expectation with respect to policy $\pi$, random status generation and transmission failure. Our objective is to find the optimal policy that minimizes the average AoI under the average transmission power constraint, which can be formulated as the following problem
\begin{prob}
	\label{p1}
	\begin{small}
	\begin{equation}
	\begin{aligned}
	& \min_{\pi} C(\pi|s_0),\\
	& {\rm{s.t.}} \    D(\pi|s_0)\leq \Gamma_{max}.
	\end{aligned}
	\end{equation}
    \end{small}
\end{prob}
%Starting from initial state $s_0$, any stationary deterministic policy makes the state return to state $(l,l,b_0)$, where $l$ is the fixed transmission times, within the maximum transmission requirement.    

Here, we assume that the IIoT device and the destination are synchronized at the beginning. That is, the initial state $s_0=(1,1,b_0)$, where $b_0$ follows status generation model $b_0\sim B(0)$. Similar to \cite{b0,b01}, we assume that the formulated problem above is always feasible and the MDP here is unichain MDP, that is, under any stationary deterministic policy $\pi \in \Pi^{MD}$, corresponding Markov chain has single (aperiodic) ergodic class \cite{b1}. As instantaneous reward in our problem satisfies the sufficient condition 
\begin{equation}
\label{e6}
\forall n \in R, {\rm the \ set} \ \{s \in S: \inf_a r(s,a) < n\}\ {\rm is \ finite}
\end{equation} to meet the growth condition \cite{b1}, placing restriction to search optimal unichain policy for feasible problem ensures the existence of optimal stationary policy according to Theorem 11.7 in \cite{b1}, which immediately leads to Corollary \ref{c1}.

\begin{corollary}
	\label{c1}
	There exists an optimal stationary policy for CMDP given in Problem \ref{p1}.
\end{corollary}
% mixture of two deterministic policy, ref(Optimal policies for controlled Markov chains with a constraint)
As we are interested in the structure of the optimal policy, we transform the formulated CMDP problem into an unconstrained MDP problem through Lagrangian relaxation as follows :
\begin{equation}
\label{e7}
L_{\lambda}(\pi|s_0)=J_{\lambda}(\pi|s_0)-\lambda\Gamma_{max},
\end{equation} where $\lambda>0$, $J_{\lambda}(\pi|s_0)=C(\pi|s_0)+\lambda D(\pi|s_0)$. The optimal policy $\pi^{*}_{\lambda^*}$ satisfies $\max_{\lambda}\min_{\pi}L_{\lambda}(\pi|s_0)$, which achieves  average AoI $C(\pi^{*}_{\lambda^*}|s_0)$ and average transmission probability $D(\pi^{*}_{\lambda^*}|s_0)$. Given a fixed $\lambda$, the following theorem shows the existence of optimal stationary and deterministic policy for the unconstrained MDP problem $\min_{\pi}J_{\lambda}(\pi|s_0)$ with countable state space and finite actions \cite{b3}\cite{b4}\cite{b5}. 
\begin{theorem}
	\label{TE}
	There exist a constant $J_{\lambda}^{*}$, a bounded function $h_{\lambda}(\delta,l,b):S \rightarrow R$ and a stationary and deterministic policy $\pi_{\lambda}^{*}$, satisfies the average reward optimality equation,
	\begin{equation}
	\label{e10}
	J_{\lambda}^{*}+h_{\lambda}(\delta,l,b)=\min_{a\in A((\delta,l,b))} (\delta +\lambda \mathbb I[a \neq 1] +{\mathbb{E}}[h_{\lambda}(\hat{\delta},\hat{l},\hat{b})])
	\end{equation} $\forall (\delta,l,b) \in S$, where $\pi_{\lambda}^{*}$ is the optimal policy, $J_{\lambda}^{*}$ is the optimal average reward, and $(\hat{\delta},\hat{l},\hat{b})$ is the next state after $(\delta,l,b)$ taking action $a$.
\end{theorem}
%
%\begin{proof}
%	The proof is omitted, due to space limit. %See the Appendix \ref{A1}.
%\end{proof}

The proof is omitted, due to space limit. Based on Theorem \ref{TE}, for the unconstrained MDP problem with fixed $\lambda$, there exists an optimal stationary and deterministic policy. Combining Theorem \ref{TE} and Theorem 4.4 in \cite{b2}, we can directly form  Corollary \ref{c2} as following,
\begin{corollary}
	\label{c2}
	If $D(\pi^{*}_{\lambda=0}|s_0) \leq \Gamma_{max}$, then there exists an optimal stationary deterministic policy for the CMDP given in Problem \ref{p1}. Otherwise, there exists an optimal stationary policy which is a randomized mixture of two stationary deterministic policies, $\pi^{*}_{\lambda_1}$ and $\pi^{*}_{\lambda_2}$, where $\Vert\lambda_{1}-\lambda_{2}\Vert\leq \epsilon$, $D(\pi^{*}_{\lambda_1}|s_0)>\Gamma_{max}>D(\pi^{*}_{\lambda_2}|s_0)$ and $\pi^{*}=\mu \pi^{*}_{\lambda_1}+(1-\mu)\pi^{*}_{\lambda_2}$, with $\mu=\frac{\Gamma_{max}-D(\pi^{*}_{\lambda_2}|s_0)}{D(\pi^{*}_{\lambda_1}|s_0)-D(\pi^{*}_{\lambda_2}|s_0)}$.
\end{corollary}
Then, Problem \ref{p1} can be solved in the following three steps:
\begin{enumerate}[Step 1:]
	\item Solve unconstrained MDP with $\lambda=0$ to judge whether $D(\pi^{*}_{\lambda=0}|s_0) \leq \Gamma_{max}$ holds. If so, $\pi^{*}=\pi^{*}_{\lambda=0}$, stop. Otherwise go to Step 2.
	\item Search $\lambda_1$ and $\lambda_2$, and solve the corresponding Lagrangian relaxed unconstrained MDP problem, where $\Vert\lambda_{1}-\lambda_{2}\Vert\leq \epsilon$ and $\small D(\pi^{*}_{\lambda_1}|s_0)>\Gamma_{max}>D(\pi^{*}_{\lambda_2}|s_0)$. 
	\item Compute the optimal stationary policy $\pi^{*}$ as a randomized mixture of $\pi^{*}_{\lambda_1}$ and $\pi^{*}_{\lambda_2}$. 
\end{enumerate}

\section{Structural Results on Optimal Policy}
In this section, we reduce action space for each state so as to reduce computation complexity, and derive two structural results of the optimal policy to gain insights into relationship between system parameters and the optimal policy. First, we establish action elimination by analyzing the property of the formulated CMDP. We then unveil the monotonicity of optimal policy in terms of the AoI $\delta$ and the total transmission times of last transmitted status update $l$.
\subsection{Action Elimination}
 The state space $S=\{(\delta,l,b)\}$ can be classified as three categories: 1) a new status update is generated, i.e., $s=(\delta,l,b=1)$; 2) no new status update is generated, but the transmission of last status update is unsuccessful while total transmission times of the status update does not exceed the maximum allowable transmission times, i.e., $s=(\delta,0<l<l_{max},b=0)$; 3) and no new status update is generated, while either last transmission is successful, i.e., $s=(\delta=l,0<l\leq l_{max},b=0)$ or no status update was transmitted at last slot, i.e., $s=(\delta,0,b=0)$ as well as total transmission times of the status update reaches $l_{max}$, i.e., $s=(\delta,l_{max},b=0)$. We provide the following proposition which helps to understand why the state space is classified as above.
 \begin{proposition}
 	\label{pro1}
 	There exists an optimal policy of the CMDP of Problem \ref{p1} that will take either action $a=1$ or $a=3$ when a new status update is generated, that is $P(a_{t}=2|b_t=1)=0$, and retransmit status update only when last transmission failed and transmission time does not exceed $l_{max}$, that is $P(a_{t}=2|a_{t-1}=0\Leftrightarrow l_t=0)=0$, $P(a_{t}=2|l_t=l_{max})=0$ and $P(a_{t}=2|l_t=\delta_t)=0$. 
 \end{proposition}

The proof of the proposition follows directly by noting that there is no point retransmitting failed stale status update when a new update is generated as the transmission failure probability is the same regardless of transmission of either new or stale status. When the status update was received at time $t$, i.e., $\delta_t=l_t$, retransmitting received status update will not reduce AoI. Moreover, considering the case that when last transmission failed, remaining idle before retransmission will not decrease transmission failure probability but leads to marginal decrease of the AoI even when the retransmission is successful \cite{b0}. In addition, new status update may be generated after the idle period and transmitting a fresh status update can lead to significant AoI drop.

According to the analysis above, the action space of the three classes of states can be reduced. For {\small$s\in S_1=\{(\delta,l,b=1)\}$, $a(s)\in \{1,3\}$}; {\small$s\in S_2=\{(\delta \neq l,0<l<l_{max},b=0)\}$}, $a(s)\in \{1,2\}$; otherwise, {\small$s \notin S_1 \cup S_2 $},  $a(s)\in \{1\}$. In the next subsection, we will apply the Proposition \ref{pro1} by modifying the transition matrix P in \eqref{e3}. Specifically, {\small$\forall s_t\notin S_1 \cup S_2$, $	P(a_t=1|s_t)=1$} (transmission is not allowed); {\small$\forall s_t\in S_1 $, $P(a_t=2|s_t)=0$} and {\small$\forall s_t\in S_2 $, $P(a_t=3|s_t)=0$}. Then, we establish the monotonicity of the optimal policy for the modified CMDP.
%{\small$\forall s_t\notin S_1 \cup S_2$}

\subsection{Effect of States on Optimal Policy}
To prove the the monotonicity of the policy in state space, we need to have the following preliminaries given in Definition \ref{d1} and Lemma \ref{l1}.
\begin{definition}
	\label{d1}
	 (Superadditive and Subadditive\cite{b5}) A multivariable function $Q(\delta,l,b,a): \Delta \times L \times B \times A \rightarrow R$ is superadditive in $(\delta,a)$ for fixed parameter $l \in L$ and $b \in B$, if for all $\delta^{+}\geq\delta^{-}$ and $a^{+}\geq a^{-}$,
	 \begin{equation}
	 \label{e8}
	 \begin{aligned}
	 	 \resizebox{1\hsize}{!}{$Q(\delta^{+},a^{+};l,b)+ Q(\delta^{-},a^{-};l,b) \geq Q(\delta^{+},a^{-};l,b)+ Q(\delta^{-},a^{+};l,b)$}
	 \end{aligned}
	 \end{equation}holds. If the reverse inequality holds, then $Q(\delta,l,b,a)$ is subadditive in $(\delta,a)$ for fixed parameter $l \in L$ and $b \in B$.
\end{definition}

\begin{lemma}
	\label{l1}
	Suppose $Q(\delta,a)$ is subadditive on $(\delta,a)$, and $\min\limits_{a\in A}Q(\delta,a)$ exists. Then
	\begin{equation}
	a^*(\delta)= \min\{a^*\in \arg\min_{a\in A}Q(\delta,a)\}
	\end{equation} is monotone nondecreasing in $\delta$.
	While $Q(l,a)$ is superadditive on $(l,a)$, and $\min\limits_{a\in A}Q(l,a)$ exists. Then
	\begin{equation}
	a^*(\l)= \min\{a^*\in \arg\min_{a\in A}Q(l,a)\}
	\end{equation} is monotone nonincreasing in $l$.
\end{lemma}

The proof is similar to the proof of Lemma 4.7.1 in \cite{b5}, and hence omitted. We establish the optimality of monotone policy by proving the state-action reward function of unconstrained MDP function: 
\begin{equation}
\label{e13}
Q_{\lambda}(\delta,l,b,a)\triangleq\delta+\lambda \mathbb I[a \neq 1] +E[h_{\lambda}(\hat{\delta},\hat{l},\hat{b})]
\end{equation} is superadditive on $(l,a)$, for fixed parameter $\delta$ and $b$, and subadditive on $(\delta,a)$, for fixed parameter $l$ and $b$. Then, the optimal policy of each state is the action that achieves the minimum value as following
\begin{equation}
\label{e14}
a_{\lambda}^*(\delta,l,b)=\arg\min_{a\in A((\delta,l,b))}Q_{\lambda}(\delta,l,b,a)
\end{equation} is monotone.

\begin{theorem}
	\label{T1}
	$Q_{\lambda}(\delta,l,b,a)$ is superadditive on $(l,a)$, for fixed parameter $\delta$ and $b$,  and subadditive on $(\delta,a)$, for fixed parameter $l$ and $b$.
\end{theorem}

%\begin{proof}
%	The proof is omitted, due to space limit. %See the Appendix \ref{A2}.
%\end{proof}
%The proof is in Appendix.
The proof is omitted, due to space limit. Hence, we can conclude from Theorem \ref{T1} and Lemma \ref{l1}, that the optimal policy for $\min_{\pi}J_{\lambda}(\pi|s_0)$ is monotone nondecreasing in state $\delta$ for fixed $l$ and $b$, and nonincreasing in state $l$, for $0<l\leq l_{max}$ when $\delta$ and $b$ is fixed. In other words, there is an optimal threshold policy based on state for the unconstrained MDP with fixed $\lambda$. This structure reduces the required memory for the IIoT device as only action shifting boundary is needed to conduct the policy.
%\vspace{-.5em}
\section{Numerical Results}
\begin{figure}[tbp]
	\vspace{-1.2em}
	\centerline{\includegraphics[height=0.21\textheight,width=0.5\textwidth]{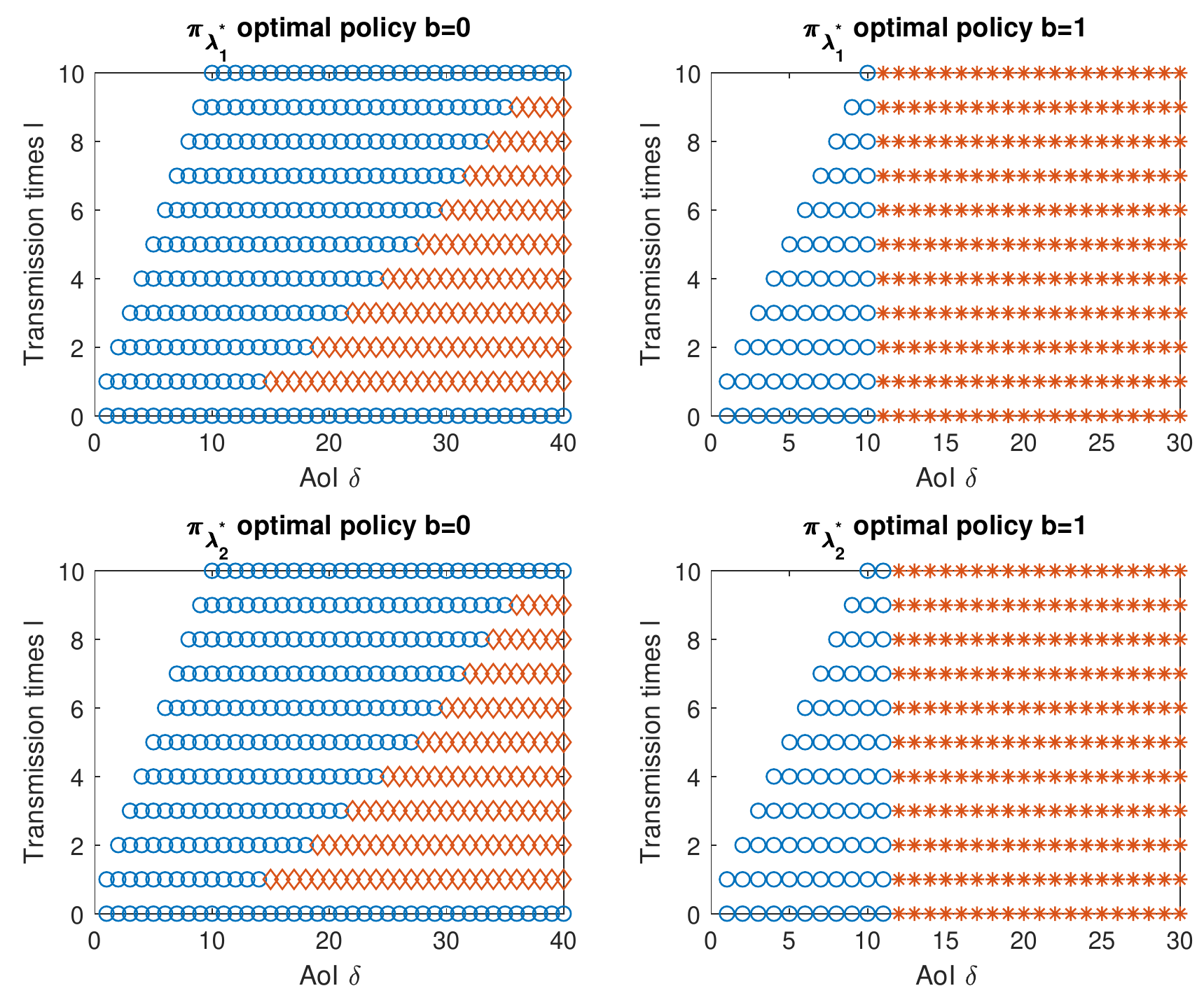}}
	\vspace{-1em}
	\caption{Structural deterministic policy for $\lambda_{1}^*$ (top) and $\lambda_{2}^*$ (bottom) where $\lambda_{1}^*<\lambda_{2}^*$, with $\Gamma_{max}=0.1, p=0.3,\gamma=0.3$. Here blue circle represents action $a=1$, diamond represents $a=2$ and star represents $a=3$.}
	\label{fig1}
	\vspace{-1em}
\end{figure}

%\vspace{-.5em}
This section provides numerical results to illustrate the analytical results provided in the preceding sections. By following the method in\cite{b6}, we apply Relative Value Iteration (RVI) on finite states ($\delta \leq 1000$ and $l_{max}= 10$) to approximate the denumerable infinite state space. We use gradient descent algorithm to calculate $\lambda_{1}$ and $\lambda_{2}$, where $\Vert\lambda_{1}-\lambda_{2}\Vert\leq 0.01$. The structure of the optimal policy for the transformed unconstrained MDP is illustrated in Fig.\ref{fig1}, which verifies that the optimal policy is monotone nondecreasing in state $\delta$ and nonincreasing in state $l$ for $0<l\leq l_{max}$. As we can see from Fig. \ref{fig1}, the threshold structure of the optimal policy is obvious. For fixed $\delta$, if $a_{\lambda}^*(\delta,l,0)=1$, $l>0$, then $a_{\lambda}^*(\delta,l+1,0)=1$. Specifically, considering the state which has transmitted the stale status for $l$ times, $l>0$, and achieves AoI $\delta$ with no new status update generated, the optimal action for the state is to remain idle (not retransmit). Then the optimal action is still to remain idle for any states that have transmitted stale status for more than $l$ times with same AoI $\delta$ and no fresh update generated. Similarly, for fixed $l$, if $a_{\lambda}^*(\delta, l, 0)=2$, then $a_{\lambda}^*(\delta+1, l, 0)=2$; if $a_{\lambda}^*(\delta, l, 1)=3$, then $a_{\lambda}^*(\delta+1, l, 1)=3$. Furthermore, when new status update is generated, the action is only determined by current AoI $\delta$. Besides, as $\lambda$ increases, the resultant policy $\pi_{\lambda}^*$ transmits less status update.
%Specifically, when no fresh status update is generated, if the optimal action for state, which has transmitted the stale status for $l$ times and achieves AoI $\delta$, is to remain idle (not retransmit), then the optimal action is still to remain idle for any states that have transmitted stale status for more than $l$ times with same AoI $\delta$.

\begin{figure}[tbp]
	\vspace{-1.1em}
	\centerline{\includegraphics[height=0.21\textheight,width=0.5\textwidth]{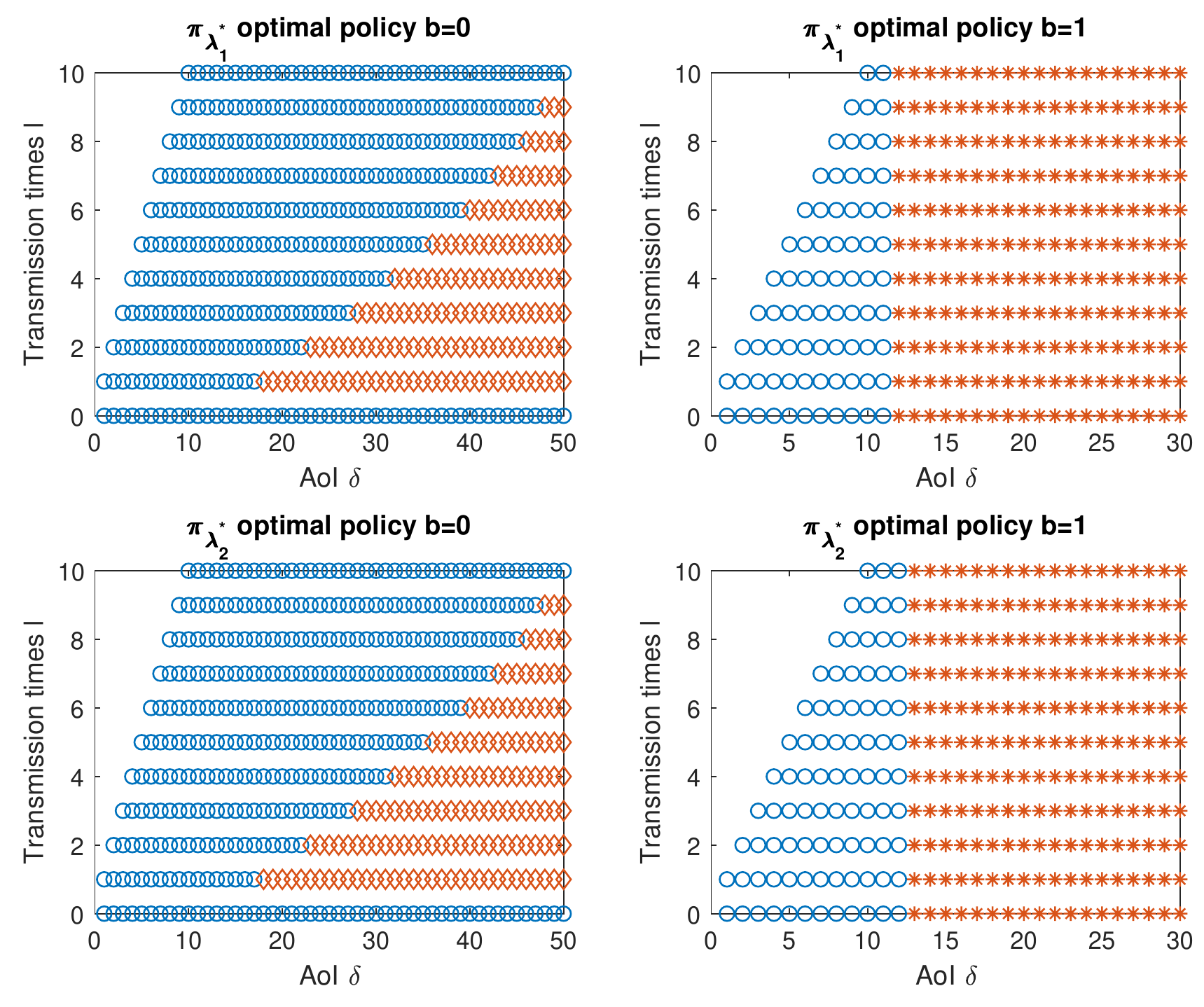}}
	\vspace{-1em}
	\caption{Structural deterministic policy for $\lambda_{1}^*$ (top) and $\lambda_{2}^*$ (bottom) where $\lambda_{1}^*<\lambda_{2}^*$, with $\Gamma_{max}=0.1, p=0.5,\gamma=0.3$. Here blue circle represents action $a=1$, diamond represents $a=2$ and star represents $a=3$.}
	\label{fig3}
	\vspace{-.8em}
\end{figure}

\begin{figure}[tbp]
%	\vspace{-1em}
	%	\includegraphics[width=0.6\textwidth]{policycmax0_1}
	\centerline{\includegraphics[height=0.21\textheight,width=0.5\textwidth]{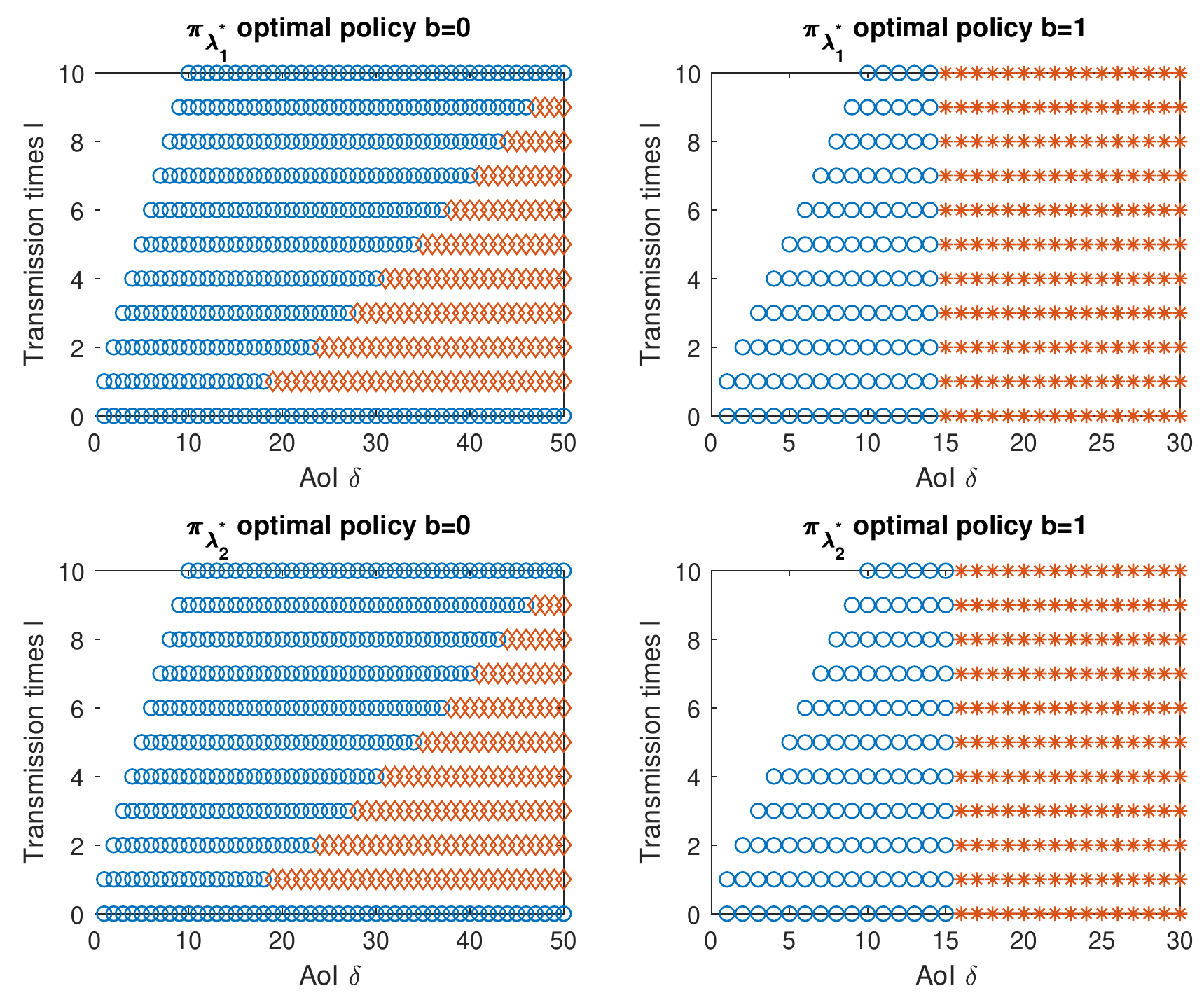}}
	\vspace{-1em}
	\caption{Structural deterministic policy for $\lambda_{1}^*$ (top) and $\lambda_{2}^*$ (bottom) where $\lambda_{1}^*<\lambda_{2}^*$, with $\Gamma_{max}=0.1, p=0.3,\gamma=0.5$. Here blue circle represents action $a=1$, diamond represents $a=2$ and star represents $a=3$.}
	\label{fig4}
	\vspace{-1em}
\end{figure}

Comparing with the structure of optimal policies of different $\gamma$ and $p$ as indicated in Fig. \ref{fig1}, Fig. \ref{fig3} and Fig. \ref{fig4}, we can see that as either transmission failure probability $\gamma$ or status generation probability $p$ increases, the action shifting boundary tends to tilt to the right for $b=0$ and shift to the right for $b=1$. As $\gamma$ increases, the state of the system are more likely to come to the case of $b=0$ with transmission failure. Hence, without action boundary tilt for $b=0$, average transmission probability will exceed $\Gamma_{max}$, which is not allowed. Moreover, for $b=1$, because of larger $\gamma$, transmitting a fresh update is less likely to reduce AoI. To balance the power limit and the AoI minimization, the action boundary of $b=1$ will move towards the right.

\begin{figure}[tbp]
	\vspace{-1em}
	\centerline{\includegraphics[height=0.21\textheight,width=0.4\textwidth]{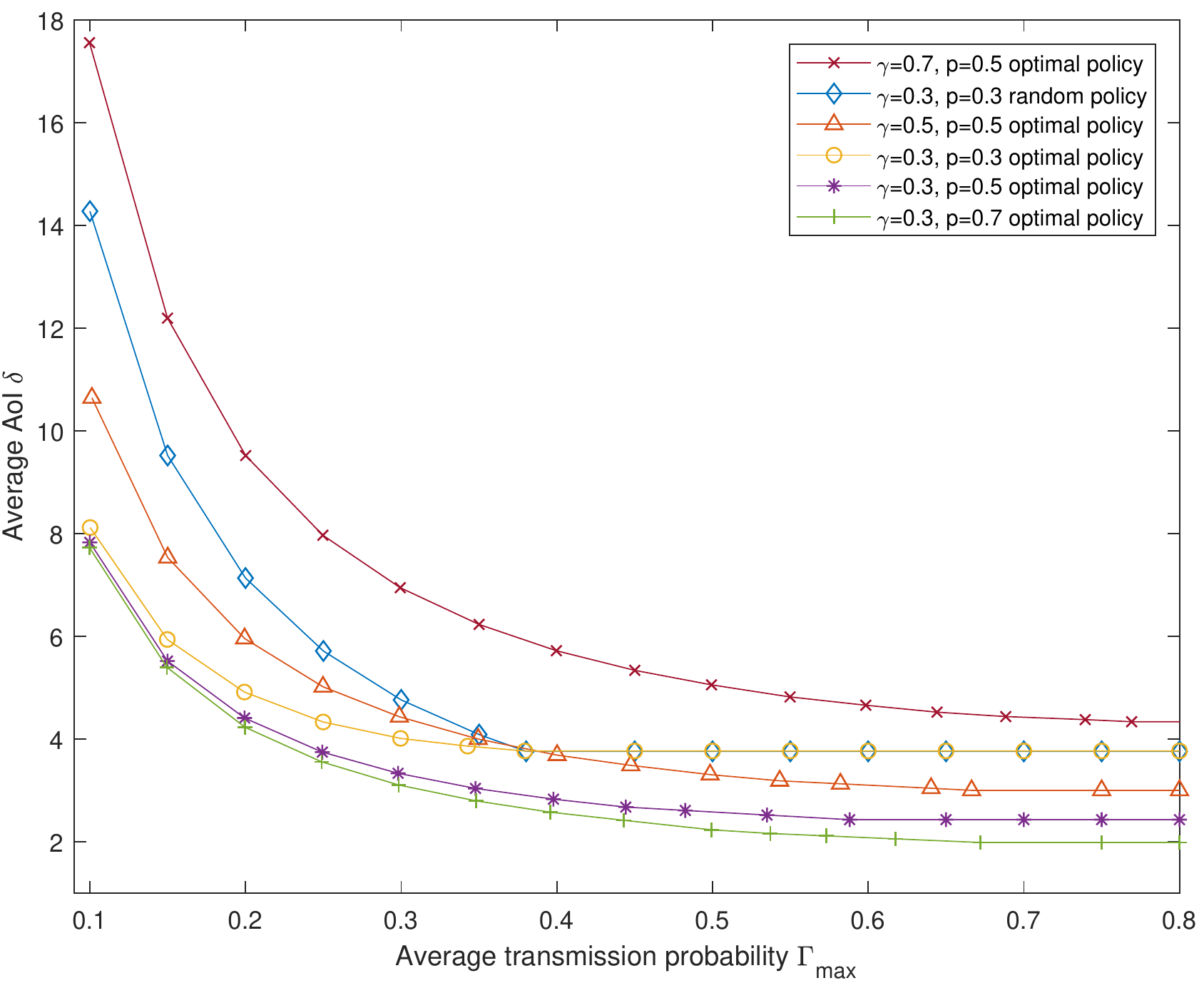}}
	\vspace{-1em}
	\caption{Tradeoff between $\Gamma_{max}$ and  expected average AoI. The result is averaged over $1000$ trials whose time horizon $T=10000$.}
	\label{fig2}
	\vspace{-.8em}
\end{figure}

Fig. \ref{fig2} illustrates the performance of the optimal policy with different transmission failure probability $\gamma$ and status generation probability $p$, comparing with a benchmarking random policy. Here, the random policy is to transmit status update with fixed probability to achieve maximum average transmission probability $\Gamma_{max}$ when a fresh status update is generated $b=1$ or no fresh status update while last transmission failed and $l<l_{max}$. As such, the transmission conditions for the random policy and optimal policy of our modified CMDP are the same, which leads to a limit on average allowable transmission probability dependent on $\gamma$ and $p$ as well as the transmission policy. When the IIoT device utilizes every chance to transmit status update, it will achieve the largest available transmission probability for fixed $\gamma$ and $p$. We can see from  Fig. \ref{fig2} that the optimal policy achieves lower average AoI, comparing to the random policy with same parameters $\gamma$ and $p$, when $\Gamma_{max}$ is small, which indicates the effectiveness of the optimal policy. As $\Gamma_{max}$ increases above certain threshold, the performance of random policy and optimal policy are the same. This is because when $\Gamma_{max}$ becomes no longer smaller than the largest available transmission probability, to achieve the lowest average AoI, both optimal policy and random policy will certainly transmit at each state when possible. Besides, the results show that larger transmission failure rate leads to larger average AoI, which is easy to understand.

%\begin{figure}[htbp]
%	\vspace{-1em}
%	\centerline{\includegraphics[height=0.21\textheight,width=0.4\textwidth]{Myfigure4}}
%	\vspace{-1em}
%	\caption{Tradeoff between $\Gamma_{max}$ and  expected average AoI. The result is averaged over $1000$ trials whose time horizon $T=10000$.}
%	\label{fig2}
%	\vspace{-.8em}
%\end{figure}

 When the status update is generated frequently (larger $p$), the average AoI decreases, and the speed of the decrease becomes slower due to the average transmission probability limit $\Gamma_{max}$, as shown in both Fig. \ref{fig2} and Table \ref{tab1}. This is because average transmission probability constraint $\Gamma_{max}$ makes it impossible to timely update each status update. When the status generation probability increases, the average transmission probability of fresh status update, $P(a=3|b=1)$, decreases, as shown in Table \ref{tab1}. The action shifting boundary movement trend observed by comparing Fig. \ref{fig3} to Fig. \ref{fig1} verifies this as well. To satisfy the average transmission power constraint, the action shifting boundary should shift to the right for $b=1$. The action boundary for $b=0$ will consequently tilt to the right. As such, the decrease of the average AoI becomes less remarkable as the status generation probability $p$ increases. In addition, when generation probability of status update equals $1$, the system simplifies into the \textit{generate at will}, and the problem becomes the ARQ protocol scheduling problem considered in \cite{b0} which is thus a special case in our model. Besides, from the intersection between optimal policy for the case $\gamma=0.3,\ p=0.3$, and optimal policy for the case $\gamma=0.5, \ p=0.5$, we can deduce that when $\Gamma_{max}$ is small, transmission failure probability plays an important role in AoI minimization and status generation probability is more crucial when $\Gamma_{max}$ is large. This is because when $\Gamma_{max}$ is small, even when the generation probability $p$ increases, fresh status update cannot be all transmitted, thus, the effect of generation probability is not significant comparing to the transmission failure probability. When $\Gamma_{max}$ is large, transmission of the status updates is no longer limited, larger  $p$ indicates more chance to update status , which make its effect on the average AoI outweigh that of larger transmission failure probability.
%Because retransmission does not decrease the transmission failure probability and the $\Gamma_{max}$ is limited, the influence of maximum transmission times on optimal policy is not significant. We do not further analyze the influence. 

\begin{table}[htbp]
	\vspace{-1em}
	\caption{Relationship among status generation probability $p$, average AoI and $P(a=3|b=1)$ when $\Gamma_{max}=0.3$ and $\gamma=0.3$.}	
	\vspace{-1.5em}	
	\begin{center}
			\begin{tabular}{*{7}{c}}
			\hline
			\ \textbf{$p$} & \ 0.3 &\ 0.4 &\ 0.5 &\ 0.6 &\ 0.7 &\ 1\\
			\hline
			\ \textbf{Minimal AoI} & \ 4.01 & \ 3.53 &\ 3.33 &\ 3.21 &\ 3.10 &\ 2.99 \\
			\hline
			\ \textbf{$P(a=3|b=1)$} & \ 0.83 & \ 0.62 &\ 0.53 &\ 0.47 &\ 0.41 &\ 0.30 \\
			\hline
		\end{tabular}
		\label{tab1}
	\end{center}
\vspace{-2em}
\end{table}

%\vspace{-1em}
\section{Conclusions}
%\vspace{-.3em}
In this paper, we have minimized the average \textit{Age of Information} (AoI) for real-time monitoring applications of Industrial IoT, where the system status is generated randomly and transmitted under average transmission power constraint over an error prone channel. We have formulated the long-term average AoI minimization problem as an infinite time horizon Constrained Markov Decision Process with average cost criterion. We then proved that the optimal stationary policy is a randomized mixture of two deterministic monotone policies. Additionally, simulations are conducted to evaluate the influence of status generation probability and transmission failure rate on optimal policy and the resultant average AoI as well as the impact of average transmission power constraint on the average AoI. As for future work, the sampling power of each status can be included as an extra part of power consumption, and the Hybrid Automatic Repeat reQuest protocol can be applied for retransmission.% instead of the adopted Automatic Repeat reQuest protocol. 

\end{document}